\documentclass[apjl]{emulateapj}
\usepackage{apjfonts}

\newcommand{\um}{${\rm \mu m}$~}
\newcommand{\mm}{${\rm \mu m}$}

\slugcomment{Accepted for publication in ApJL} 

\begin{document}

\title{The Exceptionally Large Debris Disk around $\gamma$ Ophiuchi}
\author{K. Y. L. Su\altaffilmark{1}, 
  G. H. Rieke\altaffilmark{1}, 
  K. R. Stapelfeldt\altaffilmark{2},
  P. S. Smith\altaffilmark{1},
  G. Bryden\altaffilmark{2},  
  C. H. Chen\altaffilmark{3},
  D. E. Trilling\altaffilmark{1}
  }

\altaffiltext{1}{Steward Observatory, University of Arizona, 933 N
  Cherry Ave., Tucson, AZ 85721; ksu@as.arizona.edu}
\altaffiltext{2}{JPL/Caltech, 4800 Oak Grove Drive, Pasadena, CA
  91109}
\altaffiltext{3}{Spitzer Fellow; NOAO, Tucson, AZ 85726}

\begin{abstract}

{\it Spitzer} images resolve the debris disk around $\gamma$~Ophiuchi
at both 24 and 70 \mm. The resolved images suggest a
disk radius of $\sim$520 AU at 70 \um and $\gtrsim$260 AU at 24
\mm. The images, along with a consistent fit to the spectral energy
distribution of the disk from 20 to 350 \mm, show that the primary
disk structure is inclined by $\sim$50\arcdeg~from the plane of the sky at
a position angle of 55\arcdeg$\pm$2\arcdeg.  Among a group of twelve
debris disks that have similar host star spectral types, ages and infrared
fractional luminosities, the observed sizes in the infrared and color
temperatures indicate that evolution of the debris disks is influenced
by multiple parameters in addition to the proto-planetary disk initial mass.

\end{abstract} 

\keywords{circumstellar matter -- infrared: stars -- planetary systems
-- stars: individual ($\gamma$~Oph)}

\section{Introduction}

Planetary debris disks are one of the best means to explore the
evolution of planetary systems. The {\it Spitzer} mission has made
fundamental contributions to our understanding of them,
including documenting the wide variety in properties such as amounts
of excess emission, disk size and grain properties
derived from mineralogical features (Rieke et al.~2005, Su et
al.~2006, Trilling et al.~2008, Chen et al.~2006). Two possibilities
for this variety are: 1.)~disks share a similar evolution, 
but with a wide range of initial masses (e.g., Wyatt et
al.~2007b; Kenyon \& Bromley 2008); and 2.)~there are substantial
evolutionary differences such as large collisions
that dominate the amount of debris for a period of time
(Wyatt et al.~2007a). There may also be a mixture of these
possibilities (Rieke et al.~2005; Wyatt et al.~2007b). In any case,
understanding debris disks requires determination of their range of
behavior where critical boundary conditions that might influence
these systems are held fixed.

Here we report a detailed study of the debris disk around
$\gamma$~Ophiuchi, a 184 Myr-old A0~V star (Song et al.~2001, Rieke et
al.~2005) at a distance of 29.1$\pm$0.8 pc with an infrared excess
detected in the {\it IRAS} point-source catalog (Sadakane \& Nishida
1986). Fajardo-Acosta et al.~(1997) report marginal evidence that
$\gamma$~Oph is extended at 60 \mm. In this paper, we describe imaging
with {\it Spitzer} that confirms this result by clearly resolving the
disk at both 24 and 70 \mm. Although the radii of other resolved disks
(e.g., Fomalhaut) are typically only about 150 AU, we find
that the $\gamma$~Oph disk is almost four times as large.
We combine this size measurement with the detailed spectral
energy distribution (SED) of the disk to derive its properties, and
compare them to those of a small sample of debris disks with similar
host stars.

\section{Observations and Data Reduction} 
\label{obs} 

The observations presented here are from various {\it Spitzer}
programs. The IRS spectrum of $\gamma$~Oph was previously published
based on an older pipeline reduction (S11) by Chen et al.~(2006). Here
we present the S15 reduction with updated flux non-linearity
corrections.  Observations at 24 $\mu$m were obtained at two epochs in
standard small-field photometry mode using 3 sec$\times$4 cycles at 5
sub-pixel-offset positions, resulting in a total integration of 840
sec on source for each epoch. The observation at 70 $\mu$m was done in
fine-scale mode with 3 sec$\times$1 cycle at 4 sub-pixel-offset
positions, resulting in a total integration of $\sim$100 sec on
source. A MIPS SED-mode observation was obtained with 10 sec$\times$10
cycles, and a 1\arcmin~chop distance for background subtraction.

All of the MIPS data were processed using the Data Analysis Tool
(Gordon et al.~2005) for basic reduction (e.g., dark subtraction, flat
fielding/illumination corrections), with additional processing to
minimize artifacts (see Engelbracht et al.~2007). For the 70 \um
fine-scale data, each of the on-source exposures was subtracted from
an average background composed of before and after off-source
exposures to remove the time-dependent column offsets in the
background. The final mosaics were combined with pixels half the size
of the physical pixel scale (Figure \ref{fig1}).  The MIPS SED-mode
data were reduced and calibrated as described by Lu et al.~(2008).

\begin{figure}
\figurenum{2}
\label{sed}
\plotone{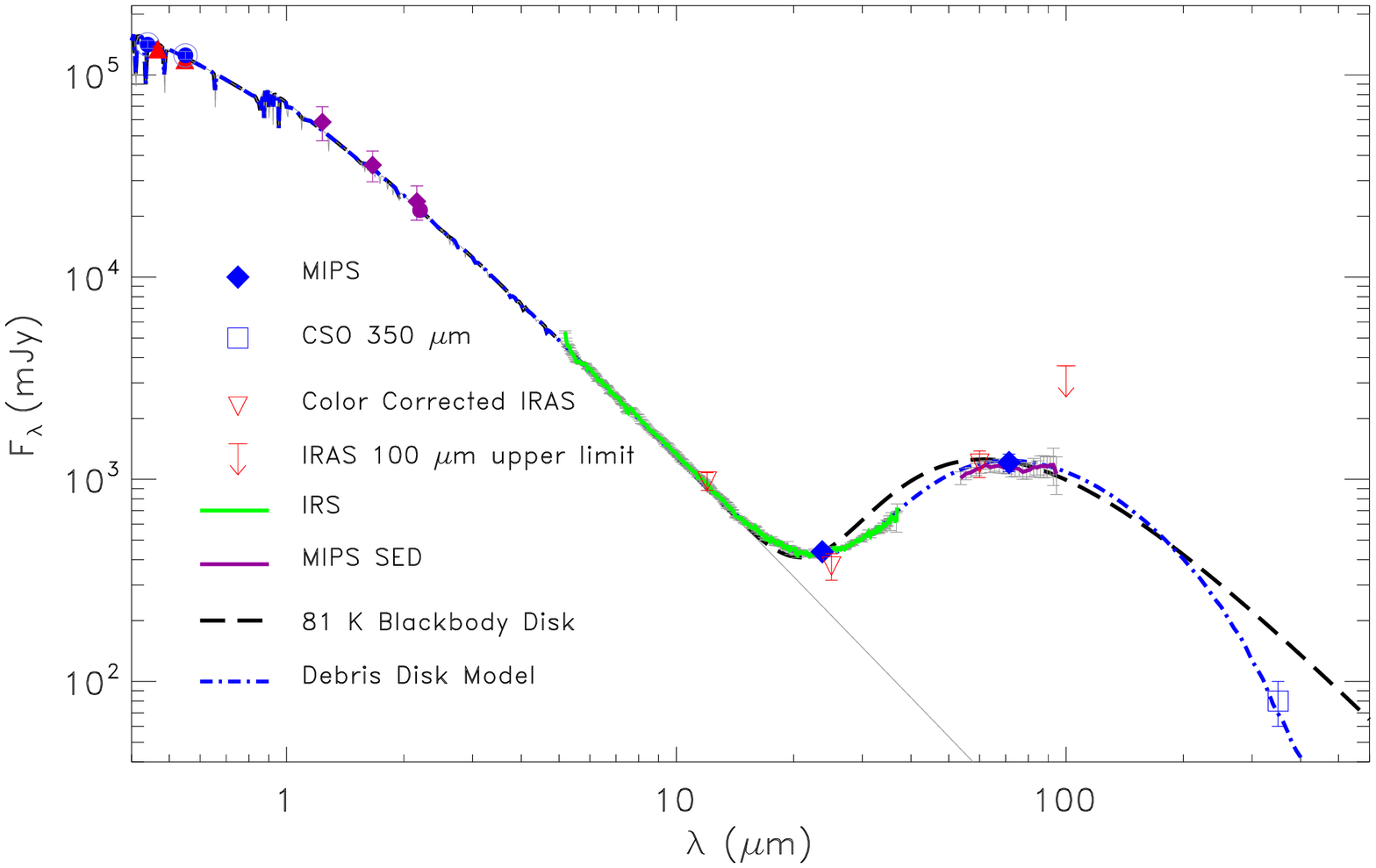}
\caption{Spectral Energy Distribution (SED) of $\gamma$~Oph. The
photometry of 80$\pm$20 mJy at 350 \um is obtained with the Caltech
Submillimeter Observatory (Dowell et al.~in prep.). The optical and
near-infrared photometry is from the Simbad database and 2MASS
catalog, while the other symbols are defined on the plot. The
dotted-dashed line is the output SED using the best-fit parameters.}

\end{figure}

At 24 \mm, the non-color-corrected measurement gives 434$\pm$3 mJy
using an aperture radius of 15\arcsec~with a background annulus of
30\arcsec--42\arcsec, and an aperture correction of 1.143. The
expected photospheric flux density in the 24 \um band is 236 mJy,
resulting in a 24 \um infrared excess of 202 mJy after a color
correction of 1.021 (for a blackbody temperature of 80 K, which is the
characteristic color temperature of the disk (see \S 3.2)). The total
color-corrected flux density in the 24 \um band is 438$\pm$18 mJy,
including the uncertainty in the calibration factor of 0.0454 MJy
sr$^{-1}$ (DN s$^{-1}$)$^{-1}$ (Engelbracht et al.~2007). For the 70
\um fine-mode photometry, we use an aperture with a radius of
16\arcsec, sky annulus of 18\arcsec--39\arcsec, and an aperture
correction of 1.9. A total flux density of 1140 mJy was estimated
using a flux calibration factor of 2894 MJy sr$^{-1}$ MIPS70F$^{-1}$
(Gordon et al.~2007). The expected photospheric flux density in the 70
\um band is 26 mJy; therefore, the disk flux density is 1185 mJy after
a color correction of 1.064 (for a blackbody temperature of 80K). The
total color-corrected flux density in the 70 \um band is 1211$\pm$121
mJy. All available photometry and spectra of $\gamma$~Oph are shown in
the SED plot in Figure \ref{sed}.

\section{Analysis} 
\subsection{PSF Subtraction} 
\label{analysis:psfsubtraction} 

The excess emission at 24 \um is roughly equal in brightness to the
expected photospheric value and yet the source appears point-like at
24 \mm. To reveal the disk, the Point Spread Function (PSF) must be
subtracted. We used deep observations of the non-excess A4 V star
$\tau_3$~Eri\footnote{MIPS 24 \um measurement (247$\pm$3 mJy) and an
updated IRS spectrum show no excess in this source.} to determine the
PSF.  At 24 \mm, the Full-Width-Half-Maximum (FWHM$_{24}$) of the
$\tau_3$~Eri image is 5\farcs52$\times$5\farcs48 using a 2-D Gaussian
fitting function on a field of 26\arcsec with an uncertainty of
0\farcs02$\times$0\farcs02 based on $\sim$100 calibration star
observations. The $\tau_3$~Eri image was scaled to the expected
photospheric flux density of $\gamma$~Oph and to 1.6 times this value
for PSF subtraction. The higher value (over-subtraction) allows for
the possibility of a significant portion of the excess emission lying
close to the star as in Fomalhaut (Stapelfeldt et al.~2004). The
results of the PSF subtractions are shown in Figure \ref{fig1}.  An
elliptical outer disk along the position angle of $\sim$56\arcdeg~is
seen in the over-subtracted image, while the photospheric-subtracted
image confirms that much of the disk emission lies within the PSF. The
disk is at least 9\arcsec ($\sim$260 AU) in radius at 24 \mm.

To test for any color dependence in this process, we subtracted the
image of $\tau_3$~Eri from the image of $\zeta$~Lep, a debris disk
known to be largely confined to within 3 AU (0\farcs14) radius at 20
\um (Moerchen et al.~2007), with various scaling factors. There are no
disk-like residuals in the subtracted image. The FWHM$_{24}$ of the
photosphere-subtracted $\zeta$~Lep image (5\farcs61$\times$5\farcs55)
gives a characteristic resolution for a red point source, slightly
broader than given by a blue PSF ($\tau_3$~Eri). The FWHM$_{24}$ of
$\gamma$~Oph before photospheric subtraction is
5\farcs84$\times$5\farcs61, which is only 1.06$\times$1.03 broader
than a blue PSF. However, this difference is significant at
16-$\sigma$ compared to a blue PSF and 11-$\sigma$ to a red PSF.  The
FWHM$_{24}$ of the $\gamma$~Oph disk after photospheric subtraction is
6\farcs41$\times$5\farcs75, which is 1.14$\times$1.03 broader than a
red PSF represented by the unresolved core of the $\zeta$~Lep disk.

The source appears slightly elongated at 70 \um (Figure 1c). No
photospheric subtraction is needed because the disk is $\sim$50 times
brighter than the stellar photosphere. However, the faint extended
disk emission is more evident after the subtraction of a PSF
representing the bright inner disk (Figure 1d). The faint extended
emission can be traced up to $\sim$18\arcsec ($\sim$520 AU, at
3-$\sigma$ level) in radius along the position angle of
$\sim$55\arcdeg. The disk has a FWHM$_{70F}$ of
18\farcs5$\times$16\farcs1 determined using a Gaussian fitting on a
field of 55\arcsec. Examining two blue calibration stars (HD 48348 and
HD 48915), a typical FWHM$_{70F}$ of 15\farcs3$\times$14\farcs4 with
an uncertainty of 0\farcs2$\times$0\farcs06 was determined, suggesting
that the $\gamma$~Oph image core at 70 \um is 1.2$\times$1.1 times
wider than a blue PSF, and resolved at 16- and 28-$\sigma$ levels for
the major and minor axises. The position angle of the disk determined
from both 24 and 70 \um images is 55\arcdeg$\pm$2\arcdeg. The MIPS
slit (width of 19\farcs8) was at the position angle of
$\sim$95\arcdeg~when the SED-mode observation was taken. A small
fraction of the faint extended disk emission was outside the slit,
which may explain the slightly lower ($\sim$5\%) SED-mode flux
compared to the fine-mode photometry.

\subsection{Spectral Energy Distribution Fitting} 

\begin{figure} \figurenum{3}
\label{excess_sed} 
\plotone{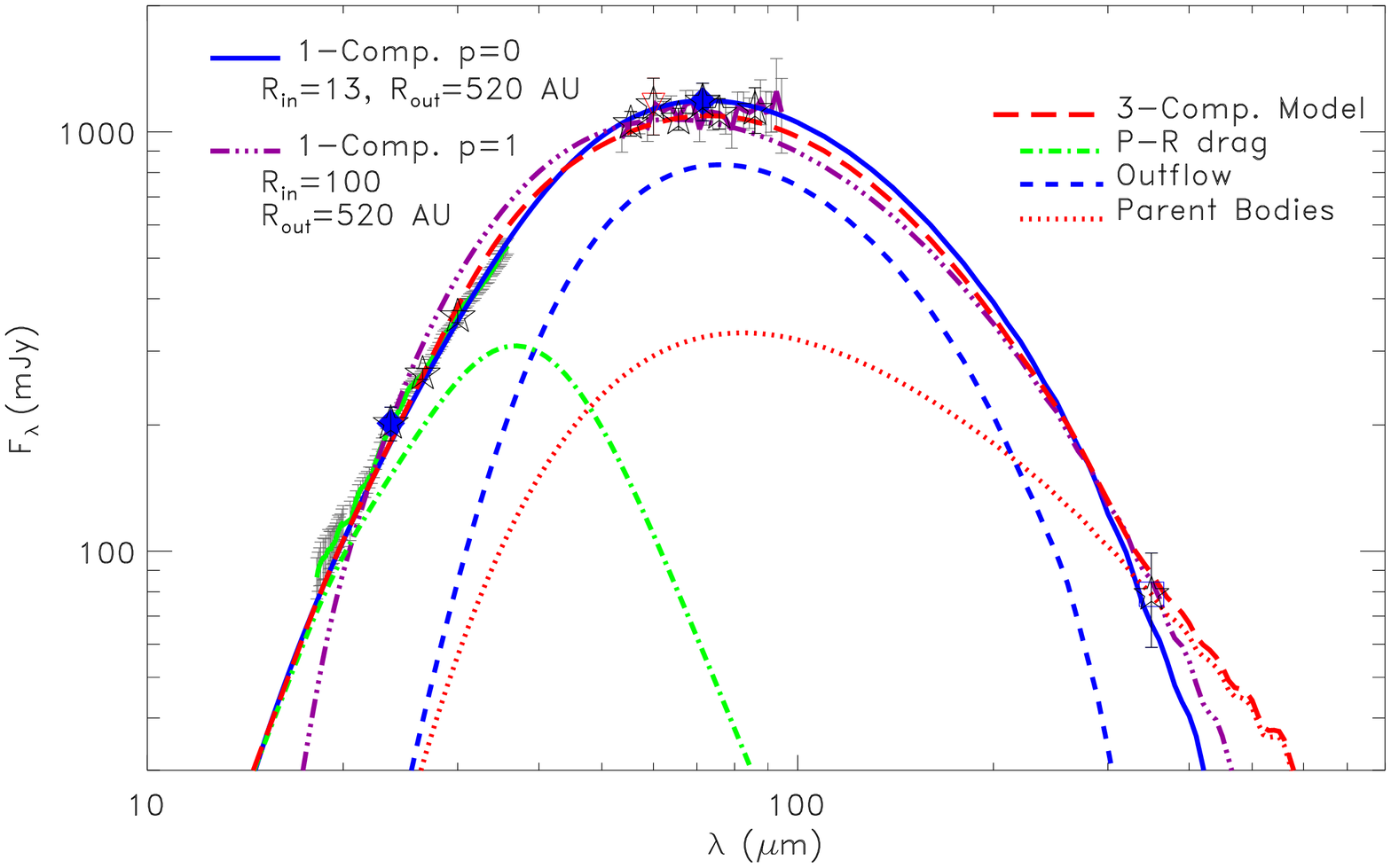}
\caption{The excess SED of the $\gamma$~Oph disk. Symbols and lines
  are the same as in Figure \ref{sed} except for the stars showing the ten
  photometric points used for $\chi^2$ fitting. Two single
  component models are shown ($p$=0 with grains of 5--63 \um and
  $\chi^2$=0.8, and $p$=1 with grains 1.6-100 \um and $\chi^2$=10.4). 
  As an illustration, the total SED of a three-component
  (not unique) disk model with $\chi^2$=0.7 is shown along with the SED of
  each component.}
\end{figure}

The broadband excess photometry at 24 and 70 \um indicates a color
temperature (T$_c$) of 81 K, which over predicts the observed flux at
28--35 \um and 55--65 \um (see Figure \ref{sed}). This suggests that
the dust grains in the system have a wider range of temperatures. To
model the disk SED (Figure \ref{excess_sed}), we use a simple
geometrically-thin debris disk model where the central star is the
only heating source and the dust is distributed radially between inner
(R$_{in}$) and outer (R$_{out}$) radii according to a $r^{-p}$ power
law for the surface number density. The grains in the disk follow a
$n(a)\sim a^{-q}$ size distribution with a minimum radius of $a_{min}$
and a maximum radius of $a_{max}$, assumed to be the same throughout
the disk. To minimize the free parameters, we assume that the grains
are astronomical silicates (Laor \& Draine 1993) with a density of 2.5
g cm$^{-3}$ and are in theoretical collisional equilibrium;
i.e. $q=-3.5$.

We determined the best-fit Kurucz model for the stellar photosphere by
fitting the optical to near-IR photometry (Su et al.~2006), and estimated the
stellar temperature and luminosity to be 9750 K and 28.6 L$_{\sun}$.
With these stellar parameters, grains smaller than
$\sim$5 \um are subject to radiation pressure blowout since $\beta$,
the ratio between radiation and gravitational forces (Burns et
al.~1979), is $>$ 0.5. To capture
the slope of the IRS spectrum and MIPS SED-mode data, we selected two
points (26.5 and 30 \mm) in the IRS spectrum and four wavelengths
(55.4, 65.6, 75.8, and 86.0 \mm) in the MIPS SED-mode data for the global
SED fitting. Combined with the broadband photometry, a total of ten data
points shown as stars in Figure \ref{excess_sed} are used to compute
$\chi^2$ to determine the best-fit debris disk model.

To fit the fairly flat slope between 55 and 95 \um and the observed
flux at 350 \um simultaneously, the maximum grain size has to be
$<$100 \mm. This might be an artifact of the assumption that the
grains have the same size distribution throughout the disk (see \S
\ref{discussion}).  In addition, since there are no obvious silicate
features, the majority of the grains in the disk are larger than a few
microns. Therefore, we assume $a_{min}$ to be the blowout size,
$\sim$5 \mm, and $a_{max}$ to be $\sim$63 \mm, and that the disk has a
constant surface density consistent with a Poynting-Robertson
(P-R) drag dominated disk. With these grain properties and density
distribution, the SED is consistent 
with a 1-component disk with an inner radius of 13$^{+4}_{-3}$ AU
and outer radius of 430$^{+130}_{-30}$AU. However, a P-R drag dominated
disk is not expected physically given the high disk density
(fractional luminosity $f_d\sim$9$\times$10$^{-5}$) since collisions
should destroy grains before they can spiral inward (Wyatt 2005).

We also explored the possibility of fitting the SED with an
outflow disk, in which the majority of the grains are smaller than the blowout
size and the surface density follows a $r^{-1}$ power law (e.g., the
Vega debris disk; Su et al.~2005). Various
combinations of $a_{min}$, $a_{max}$, R$_{in}$ and 
R$_{out}$ were tried, and none yield a correct spectral slope
between 20--30 \um (one example is shown in Figure
\ref{excess_sed}). In addition, because smaller grains have higher
temperatures than larger ones at the same radius, models including
grains smaller than the blowout size force R$_{in}$ to be a large
distance ($\sim$100 AU=3\farcs4), contradicting the
observed compact core of the disk. This suggests that the $\gamma$~Oph
system cannot be an outflow disk entirely. 

\begin{figure}
\figurenum{4}
\label{cut24n70}
\plotone{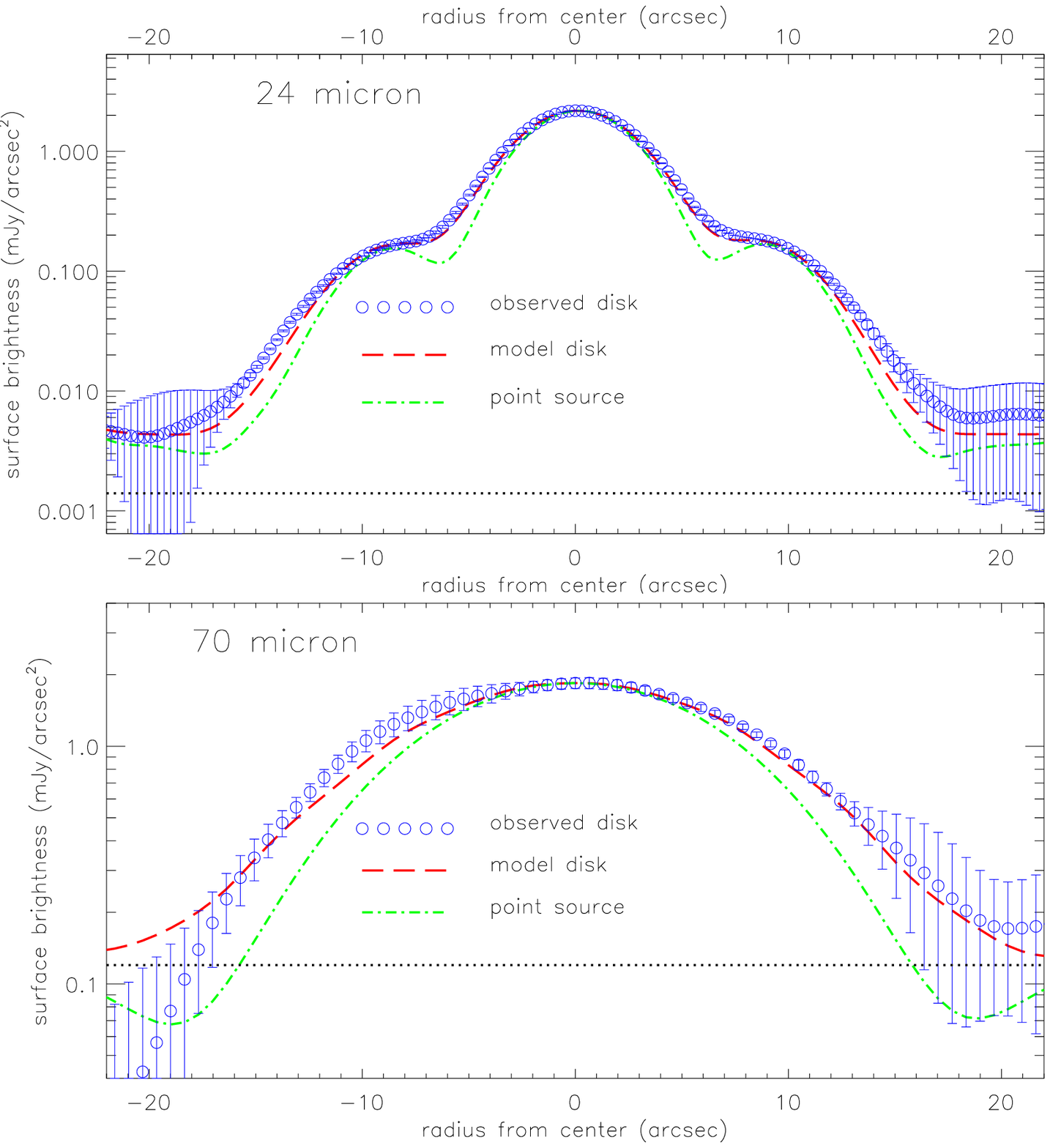}
\caption{Comparison cuts along the disk major axis at 24 \um (upper
  panel) and 70 \um (lower panel). The cuts at 24 \um are averaged
  over a width of 5\arcsec, and a width of 10\farcs5 at 70 \mm. A
  dotted line in each panel indicates a nominal 1-$\sigma$ detection on
the image.}
\end{figure}

\subsection{Debris Disk Imaging Fitting}

Limited by the spatial resolutions, no multiple disk components can be
distinguished in the observed images except for the extent of the
disk. To further constrain the model, we constructed model images with
the best-fit SED parameters ($p$=0, $a_{min}\sim$5 \mm,
$a_{max}\sim$63 \mm, R$_{in}$=13$^{+4}_{-3}$ AU, and
R$_{out}$=430$^{+130}_{-30}$ AU) from the SED fitting. These
model images were then convolved with appropriate PSFs to compare with
the observed images based on the FWHMs and radial surface brightness
distributions. 

For a geometrically thin disk model, the observed disk morphology only
depends on how the disk is viewed, i.e., the inclination angle of the
disk with respect to the plane of the sky, $i$. We first explored the
influence of the outer radius by fixing R$_{in}$=13 AU, the grain
parameters and $i$=90\arcdeg. The FWHM$_{24}$ of the convolved model
images with outer radii ranging from 350 to 1000 AU shows no change,
but a major-axis FWHM$_{70F}$ of the convolved model images varies
from 17\farcs0 to 20\farcs4 with the minor-axis FWHM$_{70F}$ of
14\farcs2. This again confirms that the 24 \um emission arises from
warm grains that are not sensitive to the outer extent of the
disk. The observed major-axis FWHM$_{70F}$ of the disk (18\farcs5)
implies a R$_{out}$ of $\sim$520 AU.  With the R$_{in}$, R$_{out}$ and
grain parameters fixed, we determined the inclination angle by fitting
both major- and minor-axis FWHMs. Convolved model images were
constructed for face-on ($i$=0\arcdeg) to edge-on ($i$=90\arcdeg)
disks in increments of 5\arcdeg. The major-axis FWHMs of the model
images remain the same within measurement errors at both 24 and 70
\mm.  From 0\arcdeg~to 90\arcdeg, the minor-axis FWHM$_{24}$ and
FWHM$_{70F}$ changed from 6\farcs1 to 5\farcs6 and 17\farcs6 to
14\farcs2, respectively.  We conclude that the observed FWHMs at 24
and 70 \um are consistent with a disk viewed at
$i$=50\arcdeg$\pm$5\arcdeg.

For a quantitative comparison, we show cuts along the disk major axis
for both the observed and best-fit model images at 24 and 70 \um 
in Figure \ref{cut24n70}. Similar cuts made on point-source PSFs,
scaled to match the peak values of the observations, are also
shown. Our simple 1-component, bound-grain 
debris disk model is in very good agreement with the observed disk
surface brightness at 24 \um over a dynamic range of two orders of
magnitude. At 70 \mm, the observed signal-to-noise ratios are not as
good as at 24 \mm, but the model agrees to within the
estimated uncertainties and approximates the data much better than a
point source.

\section{Discussion and Summary} 
\label{discussion}

Based on the {\it Spitzer} images and SED modeling, the disk around
$\gamma$~Oph has inner and outer boundaries of $\sim$13 and $\sim$520
AU, respectively, viewed at an inclination angle of $\sim$50\arcdeg,
assuming a 1-component constant-surface ($p$=0) density disk
consisting of bound (5--63 $\mu$m) astronomical silicate grains. The
dust mass is $\sim$1.0$\times$10$^{-2}$ M$_{\earth}$.  The star's
projected rotation velocity is 210 km~s$^{-1}$ (Royer et
al.~2007), suggesting an equatorial velocity of 256-300 km~s$^{-1}$
with $i$ ranging from 55\arcdeg to 45\arcdeg. Although these values
are very close to the maximum permissible velocity (Gulliver et
al.~1994), the estimated disk inclination angle is within the
permitted range of plausible stellar inclination angles.

The broadband SED alone provides too little information to constrain
properly the variables needed for even the simplest disk models, but a
better understanding can be achieved by considering other
constraints. The exact R$_{in}$ of the disk is sensitive to $a_{min}$
used in the model. The fact that the excess emission starts at
$\lesssim$15 \um suggests that dust exists as close as $\sim$10 AU,
assuming blackbody radiators. Since the grains in a debris disk are
generated by collisional cascades from large ($\sim$km-size) parent
bodies, $a_{max}$ is usually assumed to be $\sim$1000 \mm. To make the
single component model consistent with the faint extended disk
detected at 70 \um and with the assumed grain size distribution,
$a_{max}$ has to be reduced to $\sim$63 \mm. A more physically
realistic model will involve multiple components while requiring the
presence of very large grains in the disk.

We used a three-component disk to derive a model consistent with the
expected grain physics and the observed spectrum/SED. The three
components are: (1) a narrow ($\delta$R$\sim$0.1R) ring of parent
bodies at $\sim$80 AU where large grains are located; (2) an outflow
disk (80--520 AU) with density $\propto r^{-1}$ outside the narrow
ring consisting of grains being ejected by radiation pressure with
sizes smaller than the blowout size; (3) a component with density
$\propto r^0$ interior to the narrow ring and extending to $\sim$10 AU
(e.g., as of grains slowly drifting inward due to P-R drag). Many
combinations of parameters in this three-component disk model would
yield acceptable fits to the observed data. One example is shown in
Figure \ref{excess_sed}, but it is not unique without further
constraints.

Given the exceptionally large disk seen in the $\gamma$~Oph images,
one may wonder whether it is another Vega-like disk (an outflow disk
observed at 24 and 70 \um extending from a ring of large parent bodies
that dominate the emission at 850 \mm, Su et al.~2005). The failure to
fit all of the available data with our simple 1-component outflow
model suggests that the contribution by grains escaping from the
$\gamma$~Oph disk is weak, unlike for the Vega disk where such grains
dominate the disk emission ($\sim$100\% at 24 \um and $\sim$75\% at 70
\mm). Persuasive evidence of $\gamma$~Oph being another Vega-like disk
would require the disk size at 350 \um to be substantially smaller
than the size inferred in the infrared. However, the signal-to-noise
of the 350 \um data is too low to show the extent of the
emission. Observations at longer submillimeter or millimeter
wavelengths and higher spatial resolution are needed to determine the
exact location and amount of the cold component.

An attractive comparison sample of debris disks for $\gamma$~Oph is
defined by systems with stellar ages between 150 and 400 Myr, spectral
types between A0V and A3V, and $f_d$ between $10^{-5}$ and $10^{-4}$,
for which Su et al.~(2006) find a total of 12 stars\footnote{Besides
$\gamma$~Oph, Vega and Fomalhaut: $\iota$~Cen: T$_c\sim$250 K,
$\zeta$~Lep: T$_c\sim$206 K, $\alpha$~CrB: T$_c\sim$125 K,
$\beta$~UMa: T$_c\sim$116 K, 30~Mon: T$_c\sim$105 K, HD~38056:
T$_c\sim$96 K, HD~79108 : T$_c\sim$88 K, $\lambda$~Boo: T$_c\sim$86 K,
$\gamma$~Tri: T$_c\sim$75 K.}  including the resolved disks around
Vega and Fomalhaut. The disk around Fomalhaut defines the expected
template structure given our knowledge of the solar system. It
consists of two separate dust belts: a Kuiper-Belt-analog ring with a
radius of $\sim$150 AU prominent at $\lambda \gtrsim$70 \um and a fair
amount of zodiacal-analog warm dust located $<$20 AU from the
star. Little emission originates from outside of the Kuiper Belt
ring. In the Fomalhaut disk, the ratio between the central unresolved
and the extended emission is 3--6 at 24 \um (Stapelfeldt et al.~2004),
while the ratio is $\sim$2--3 in the $\gamma$~Oph disk. The
Kuiper-Belt-analog ring in the Fomalhaut system totally dominates the
emission at 70 \mm, but at the distance of $\gamma$~Oph
($\sim$4$\times$ that of Fomalhaut), such a ring would be unresolved.

Two of the twelve systems ($\zeta$~Lep and $\iota$~Cen) have 24 to 70
\um color temperatures $>$ 200 K (Su et al.~2006). Ground-based 20 \um
imaging of $\zeta$~Lep shows it to be dominated by a very compact
inner warm disk (Moerchen et al.~2007).  Three more systems (30 Mon,
$\beta$~UMa, $\alpha$~CrB) have color temperatures of 100--150 K,
indicating relatively large amounts of material in their inner zones.
All of these systems are likely to have more debris-generating
activity close to the star than the disks around Fomalhaut (color
temperature of 70 K) or $\gamma$~Oph. Of the remaining seven systems,
at least two have anomalous activity in their outer zones: Vega with
its large ($\sim$800 AU) halo, and $\gamma$~Oph with a disk $\sim$520
AU in radius.

It appears that much of the overall debris disk behavior can be
explained as the result of similar evolution from starting points that
differ in the initial disk mass (Wyatt et al.~2007b).  However, the
large variety in behavior among the sample of twelve middle-aged disks
demonstrates that one or more additional parameters must also be at
play. For instance, the presence and number of massive planets, the
recent collisional history of the system (Kenyon \& Bromley 2005,
2006), as well as the variety in the initial disk size may all play
important roles in the wealth of disk behavior.

\acknowledgments
Based on observations made with the {\it Spitzer Space Telescope},
which is operated by the Jet Propulsion Laboratory, California
Institute of Technology. Support for this work was provided by NASA
through contract 1255094 and 1256424 issued by JPL/Caltech to the
University of Arizona.


 \begin{figure*}
 \figurenum{1}
 \label{fig1}
 \plotone{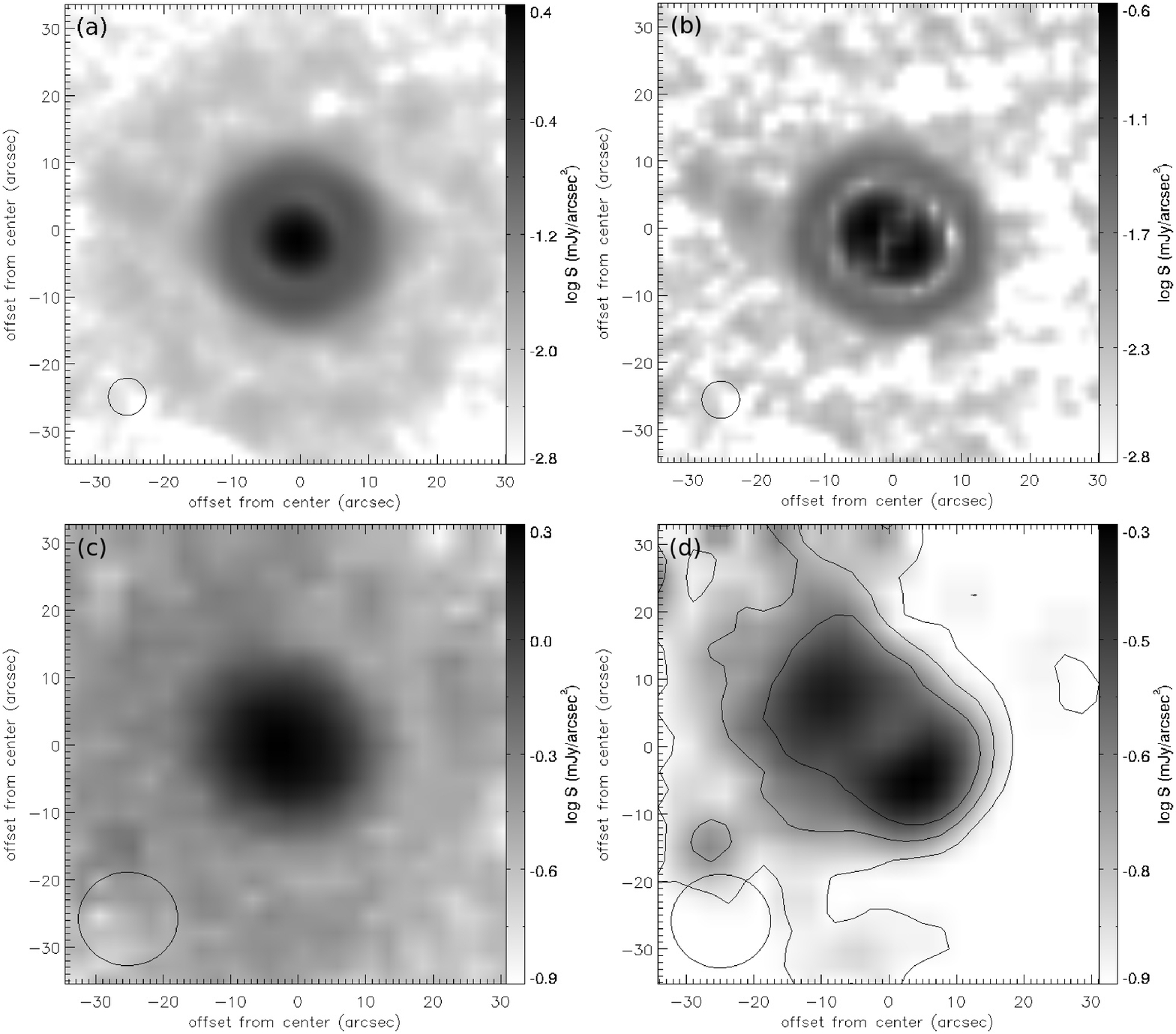} 
 \caption{MIPS 24 and 70 \um images of $\gamma$~Oph with colorbars
   showing the surface brightness levels between the 1-$\sigma$
   detection and the peak value of the image. (a) shows the deep 
   24 \um image of $\gamma$~Oph after photospheric PSF 
   subtraction. (b) shows the over-subtracted 
   image of $\gamma$~Oph at 24 \mm, where an elliptical disk is evident. (c)
   shows the disk image at 70 \mm. 
   (d) shows the over-subtracted image at
   70 \um after applying boxcar smoothing. The contours show the
   detection boundaries at 1-, 3- and 5-$\sigma$ levels;
   1-$\sigma=$0.126 mJy arcsec$^{-2}$. All
   panels have the same orientation with N up and E to the left.
   A circle at the corner of each panel indicates the typical resolution.}
 \end{figure*}

\end{document}